\documentclass[twocolumn]{aastex631}
\usepackage{amsmath}
\usepackage{amssymb}
\usepackage{amsfonts}
\usepackage{graphicx}
\usepackage{multirow}

\begin{document}
\shortauthors{Panda et al.}
\shorttitle{Low-frequency, wideband study of FRB 20240114A with the GMRT}
\title{Low-frequency, wideband study of an active repeater, FRB 20240114A, with the GMRT}

\correspondingauthor{Ujjwal Panda}
\email{ujjwalpanda97@gmail.com, upanda@ncra.tifr.res.in}

\author[0000-0002-2441-4174]{U. Panda}
\author[0000-0002-2892-8025]{J. Roy}
\author[0000-0003-0669-873X]{S. Bhattacharyya}
\author[0009-0005-4130-892X]{C. Dudeja}
\author[0000-0002-6631-1077]{S. Kudale}
\affiliation{
  National Centre for Radio Astrophysics,
  Tata Institute of Fundamental Research,
  Pune,
  Maharashtra,
  India
  (PIN: 411007)
}

\begin{abstract}

  We report the detection of 167 bursts from an active repeater, FRB 20240114A, using the uGMRT. The observations were carried over a frequency range of 300 $-$ 750 MHz, and on 4 different dates over a period of 6 months. Our analysis indicates that the FRB's emission properties are evolving, with both the median flux and emission rate of the bursts decreasing over time. The properties of the bursts also vary widely, with a wide range of intrinsic widths (0.246 to 39.364 ms), scattering timescales (0.004 to 28.289 ms), dispersion measures (524.07 to 533.56 pc cm$^{-3}$), and band occupancies (9 to 180 MHz). The distribution for the waiting time, $t_{\mathrm{wait}}$, is bi-modal, and follows a Weibull distribution, with a shape parameter $k = 0.33$; however, when the two distributions are fit separately, they follow a Weibull distribution with $k = 0.63$ for $t_{\mathrm{wait}} < 1$ s, and a log-normal distribution with a width of $\sigma = 1.28$ for $t_{\mathrm{wait}} > 1$ s. The isotropic energy distribution is seen to follow a log-normal distribution as well, with a width of $\sigma = 0.83$.

\end{abstract}

\section{Introduction}

Discovered serendipitously in 2007 \citep{lorimer_bright_2007}, Fast Radio Bursts (FRBs) are transient sources in the radio sky. Their emission is bright, with energies ranging from $10^{35}$ to $10^{43}$ ergs \citep{zhang_energy_2021}; wideband, with bursts detected from 110 MHz \citep{pleunis_lofar_2021} to 8 GHz \citep{gajjar_highest_2018}; coherent, with brightness temperatures as high as $T_{\mathrm{B}} = 10^{36}$ K essentially eliminating the plausibility of incoherent emission mechanisms \citep{zhang_physics_2022}; and short in duration, having been detected at sub-microsecond \citep{hewitt_dense_2023, majid_bright_2021} to millisecond timescales. Current estimates of their sky rate indicate that they are ubiquitous; for example, \cite{amiri_first_2021} (and the associated erratum, \cite{collaboration_erratum_2023}) estimate a sky rate of $525 \pm 30 \, (\text{stat.}) ^{+142}_{-131} \, (\text{sys.}) \text{ bursts sky}^{-1} \text{ day}^{-1}$, above a fluence limit of 5 Jy ms at 600 MHz. The unusually high values of their dispersion measure (DM) indicate that a majority of them are extragalactic sources, which has since been confirmed via localisations and subsequent host galaxy associations (e.g.: \cite{chatterjee_direct_2017, tendulkar_host_2017, bannister_single_2019, xu_fast_2022} and many others). As per the latest version of the FRB catalogue (accessed from the Transient Name Server (TNS): \url{https://www.wis-tns.org}), 803 such events have been detected. While most of these are one-off events, a minor fraction (56 out of 803, or $\sim 7\%$) has been seen to repeat, with the first such event detected in 2016 \citep{spitler_repeating_2016}; these are called repeaters. Among these repeaters, an even smaller fraction (6 out of 56, or $\sim 11 \%$) has demonstrated hyperactivity. To date, 6 FRBs have shown such hyperactivity: FRB 20121102A, FRB 20180916B, FRB 20190520B, FRB 20201124A, FRB 20220912A, and FRB 20240114A. Observing such FRBs in their hyperactive phases provides a large sample of bursts. Individual bursts can be used to probe the complex spectro-temporal polarimetric properties of the source emission, while studying how burst parameters are statistically distributed can provide clues into the emission mechanism itself, and can lead to stronger associations with plausible progenitors.

\begin{figure*}
  \includegraphics[width=0.95\textwidth]{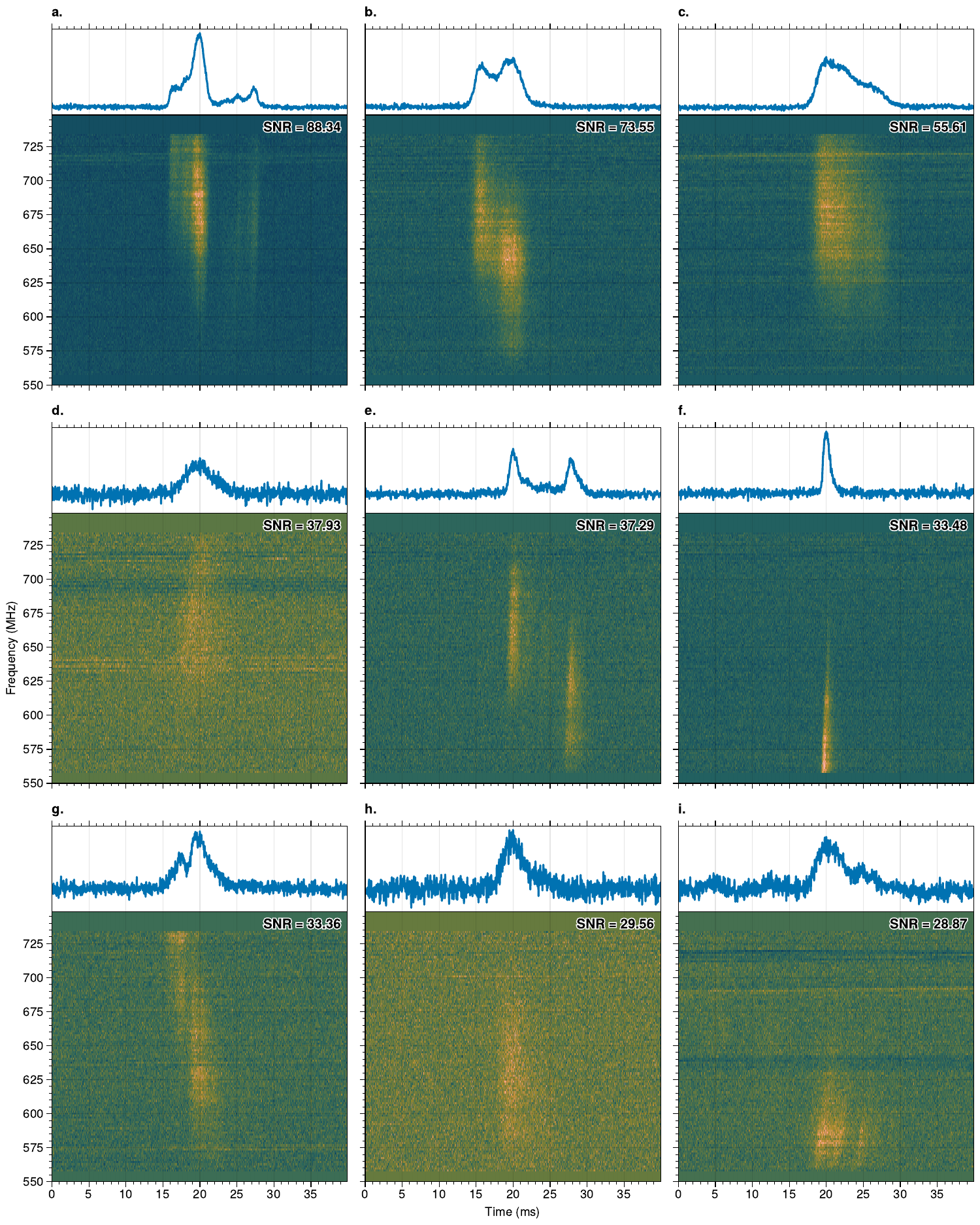}
  \caption{The dedispersed dyanmic spectra for the nine brightest bursts detected in our observations. The top panels show the dedispersed profiles for each of the bursts. All of the above bursts were detected in Band 4 (550 $-$ 750 MHz). The bursts show a variety of burst morphologies, including: band-limited emission (for example, \textbf{(d)} \textbf{(f)}, \textbf{(h)} and \textbf{(i)}), multiple components (for example, \textbf{(a)}, \textbf{(b)} \textbf{(c)}, \textbf{(e)}, \textbf{(g)}, and \textbf{(i)}), and sub-burst frequency drifting a.k.a. the \textit{sad trombone effect} (for example, \textbf{(b)}, \textbf{(e)}, and \textbf{(g)}).}\label{fig:gallery}
\end{figure*}

The CHIME/FRB collaboration reported the discovery of an active repeater, FRB 20240114A, on 26th January 2024 \citep{shin_chimefrb_2024, shin_chimefrb_2025}. A high burst rate was inferred since the source was observed at a declination of $\sim 4.4$ degrees, where the median daily exposure of the CHIME telescope is merely 4 minutes (compared to a "typical" exposure time of 15 minutes for sources near the zenith; e.g.: see Fig. 5 in \cite{amiri_first_2021}); yet, 3 bursts were detected within a span of 2 weeks. Subsequent follow-ups by multiple telescopes confirmed this hypothesis. These include Parkes/Murriyang \citep{uttarkar_detection_2024}, Westerbork \citep{ould-boukattine_bright_2024}, FAST \cite{zhang_detection_2024-1, zhang_detection_2024}, Northern Cross \citep{pelliciari_detection_2024}, MeerKAT \citep{tian_detection_2024}, uGMRT \citep{kumar_detection_2024, panda_detection_2024}, Nancay \citep{hewitt_detection_2024}, Allen \citep{joshi_wideband_2024}, Effelsberg \citep{limaye_broadband_2024}, and others \citep{ould-boukattine_over_2024}. The burst was first potentially associated with one of several galaxy clusters using DESI legacy imaging \citep{oconnor_frb_2024}, and then localized by MeerKAT \citep{tian_detection_2024} to within an accuracy of $\sim 1.5$ arcseconds, and by EVN PRECISE \citep{snelders_evn_2024} to an accuracy of $\pm 200 \text{ milliarcseconds}$. Both localizations agree within $1\sigma$ of one another, and indicate that the source is associated with J212739.84$+$041945.8, a galaxy catalogued in the Sloan Digital Sky Survey (SDSS) with a photometric redshift of $z = 0.42$ \citep{alam_eleventh_2015}. \cite{bhardwaj_redshift_2024} used the Optical System for Imaging and low-Intermediate-Resolution Integrated Spectroscopy (OSIRIS) spectrograph at the GTC telescope, and gave a spectroscopic redshift of $z = 0.13 \pm 0.0002$ for the putative host galaxy, which corresponds to a luminosity distance of $D_{L} = 630.72 \text{ Mpc}$ (using cosmological parameters given in \cite{aghanim_planck_2020}). While \cite{xing_coincident_2024} reported the presence of coincident $\gamma$-ray emission from the direction of this source using public Fermi LAT data, this was swiftly contradicted by \cite{principe_frb_2024}, who reported a non-detection. \cite{xing_detection_2024} reported the detection of a gamma-ray flare coincident with the FRB at the position determined by the EVN PRECISE collaboration \citep{snelders_evn_2024}. Multi-wavelength observation in other bands, such as in the X-ray, have not yielded positive results \citep{eppel_constraints_2025}. A persistent radio source was initially identified by \cite{zhang_discovery_2024} at L-band using public MeerKAT data, and was detected again using the uGMRT at 650 MHz \citep{bhusare_detection_2024}. The association of the PRS with the source was finally confirmed with the VLBA \citep{bruni_discovery_2024} to within the $\pm 200$ milliarcseconds position given by the EVN PRECISE collaboration \citep{snelders_evn_2024}.

In this paper, the detection of multiple bursts from FRB 20240114A with the uGMRT at low frequencies is reported, along with the results from the investigation of various statistical properties of these bursts. In \S\ref{section:observations}, the observations are described in detail, while the data analysis methodology is discussed in \S\ref{section:analysis}. The results are presented in \S\ref{section:results}, followed by a summary in \S\ref{section:summary}.

\section{Observations}\label{section:observations}

The source was observed using the uGMRT on 4 separate occasions: 25/02/24, 14/03/24, 03/07/24, and 13/07/24. Table \ref{table:obsdetails} summarises the details of all observations. A total of 22 epochs were recorded, resulting in approximately 18.63 hours of on-source time. Except for the last two observations, two beams were recorded in parallel: a phased array beam without dispersion correction (viz. PA), and a coherently dedispersed phased array beam (viz. CD). The latter removes the intra-channel dispersion smearing from each channel. This enables data recording with fewer frequency channels, thereby achieving higher time resolution without losing signal features to dispersive smearing. The observations taken on 14/03/24 used uGMRT's subarray mode, enabling simultaneous observations in Bands 3 (300 $-$ 500 MHz) and 4 (550 $-$ 750 MHz), using 10 and 12 antennas, respectively. While this lead to reduced sensitivity ($\sim 3.8$ K/Jy in both bands), it also opened up the possibility of detecting the same burst(s) in both bands. Visibilities were also recorded for the observations on 25/02/24 and 14/03/24, with the aim of constraining any continuum emission associated with the source.

All epochs were searched for single pulses using the same procedure. The data was first cleaned using the radio frequency interference (RFI) mitigation procedures in the GMRT Pulsar Tool, or \href{https://github.com/chowdhuryaditya/gptool}{\texttt{gptool}}\footnote{\url{https://github.com/chowdhuryaditya/gptool}}. Then, the data was incoherently dedispersed using \href{https://github.com/scottransom/presto}{\texttt{PRESTO}}'s \texttt{prepsubband} utility \footnote{\url{https://github.com/scottransom/presto}}, over a DM range of 520 to 535 pc cm$^{-3}$ with a DM step of 0.1 pc cm$^{-3}$, while downsampling the data to a time resolution of 327.68 $\mu$s. The DM step was chosen so as to place an adequate number of DM trials between the start and end values, which, along with downsampling, allows us to detect a larger number of bursts.. The resultant time series were then searching using \texttt{PRESTO}'s single pulse search routine, with a signal-to-noise ratio (S/N) threshold of 5. Since the number of candidates obtained was quite large, a higher S/N cutoff of 7 was placed, and the candidates were divided into low S/N (S/N $<$ 10) and high S/N (S/N $\geq$ 10) classes. In order to better judge between actual and spurious candidates, we used \href{https://github.com/astrogewgaw/candies}{\texttt{candies}}\footnote{\url{https://github.com/astrogewgaw/candies}}, which is a program used to create meaningful features from single pulse candidates to enable improved ML/DL-based classification (Panda et al. in prep). For each candidate, \texttt{candies} creates two features: the dedispersed dynamic spectrum, and the DM transform. These are then sifted through by eye. A total of 167 unique bursts were detected in all observations using the above methodology (see Table \ref{table:obsdetails}). No simultaneous bursts were detected in Bands 3 and 4.

\begin{deluxetable*}{ccccccccc}
\tablecaption{Observation details\label{table:obsdetails}}
\tablehead{\colhead{UTC Date and Time} & \colhead{Band} & \colhead{Mode} & \colhead{No{.} of channels} & \colhead{Bandwidth} & \colhead{Sampling time} & \colhead{On-source time} & \colhead{No{.} of bursts} & \colhead{Burst rate} \\ \colhead{} & \colhead{} & \colhead{} & \colhead{} & \colhead{(MHz)} & \colhead{($\mu$s)} & \colhead{(hr)} & \colhead{} & \colhead{(hr$^{-1}$)}}
\startdata
\hline
\multirow{2}{*}{\shortstack{25/02/24  \\ 03:11:30 $-$ 09:41:30}}
          & \multirow{2}{*}{\shortstack{Band 4 \\ (550 $-$ 750 MHz)}} & PA & 4096 & \multirow{2}{*}{200} &  81.92 & \multirow{2}{*}{6.50} & \multirow{2}{*}{74} & \multirow{2}{*}{11.38} \\
          &                                           & CD &  128 &                     &    2.56 &                       &                     &                       \\
\hline
\multirow{4}{*}{\shortstack{14/03/24 \\ 00:42:11 $-$ 03:56:38}}
          & \multirow{2}{*}{\shortstack{Band 4 \\ (550 $-$ 750 MHz)}} & PA & 4096 & \multirow{2}{*}{200} & 163.84 &  \multirow{2}{*}{3.24} &  \multirow{2}{*}{45} & \multirow{2}{*}{13.88} \\
          &                                           & CD &  512 &                      &  10.24 &                        &                      &                       \\
          \cline{2-9}
          & \multirow{2}{*}{\shortstack{Band 3 \\ (300 $-$ 500 MHz)}} & PA & 4096 & \multirow{2}{*}{200} & 163.83 & \multirow{2}{*}{3.24} &  \multirow{2}{*}{20} & \multirow{2}{*}{6.17} \\
          &                                           & CD &  512 &                      &  10.24 &                       &                      &                      \\
\hline
\multirow{2}{*}{\shortstack{02/07/24 $-$ 03/07/24 \\ 21:24:27 $-$ 02:17:03}}  & \multirow{2}{*}{\shortstack{Band 4 \\ (550 $-$ 750 MHz)}} & \multirow{2}{*}{PA} & \multirow{2}{*}{4096} & \multirow{2}{*}{200} & \multirow{2}{*}{163.84} & \multirow{2}{*}{4.88} &  \multirow{2}{*}{19} &  \multirow{2}{*}{3.89} \\
                                        &                          &    &      &     &        &       &     &       \\
\hline
\multirow{2}{*}{\shortstack{12/07/24 \\ 20:14:53 $-$ 21:00:50}}  & \multirow{2}{*}{\shortstack{Band 4 \\ (550 $-$ 750 MHz)}} & \multirow{2}{*}{PA} & \multirow{2}{*}{4096} & \multirow{2}{*}{200} & \multirow{2}{*}{163.84} &  \multirow{2}{*}{0.77} &   \multirow{2}{*}{9} & \multirow{2}{*}{11.75} \\
                                        &                          &    &      &     &        &       &     &       \\
\hline
\textbf{Total}     &                          &    &      &     &        & \textbf{18.63} & \textbf{167} & \textbf{10.85} \\
\enddata
\end{deluxetable*}

\begin{figure*}
  \begin{center}
    \includegraphics[width=\textwidth]{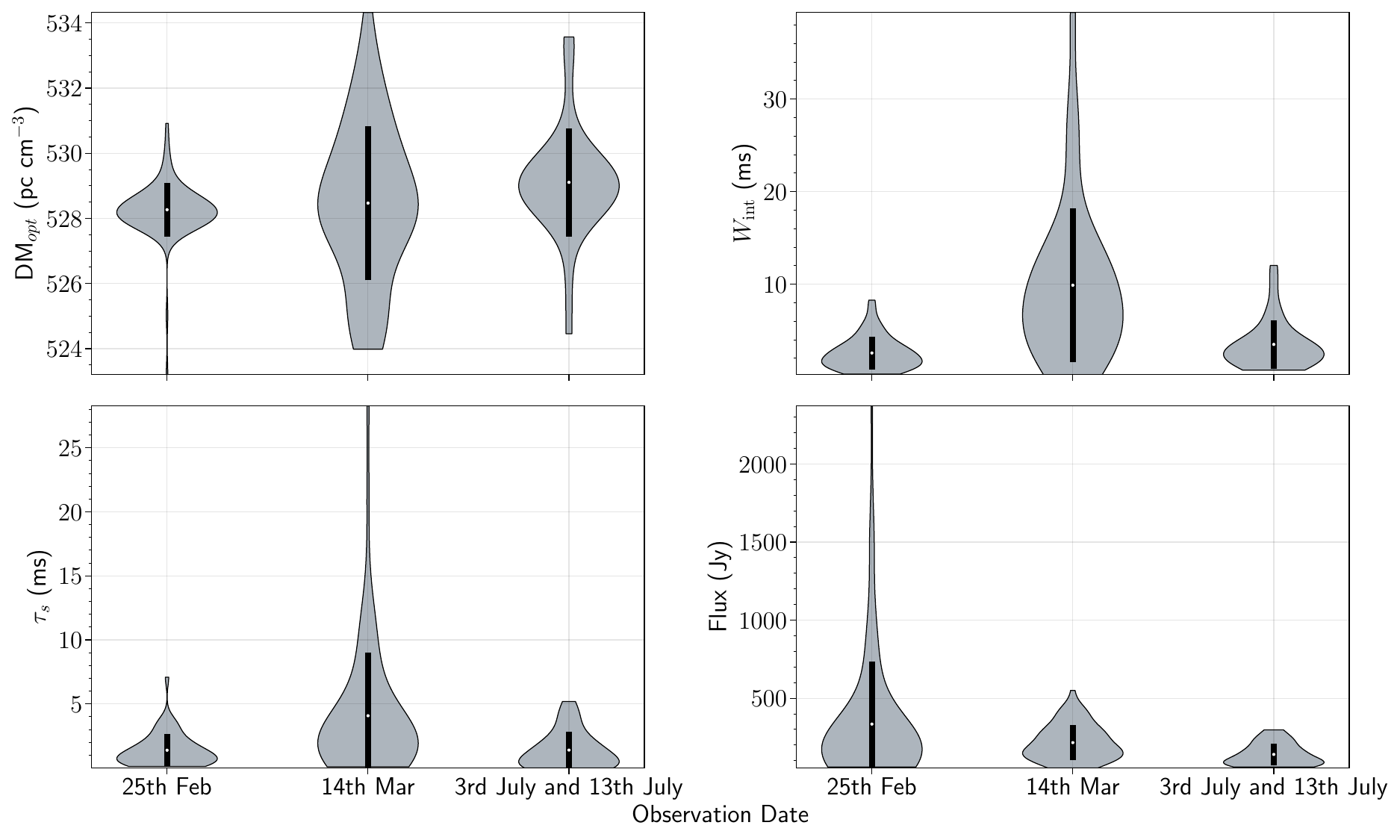}
  \end{center}
  \caption{Violin plots for the distributions of \textbf{(a)} the structure-optimised $\mathrm{DM}$, \textbf{(b)} the intrinsic width, $W_{\mathrm{int}}$, \textbf{(c)} the scattering timescale, $\tau_{s}$, and \textbf{(d)} the flux of the detected bursts on each of the 4 separate dates when we observed the source.}\label{fig:distributions}
\end{figure*}

\section{Data Analysis}\label{section:analysis}

The bursts were first chopped out from the original filterbank files for each epoch, and stored as individual filterbank files. Coherently dedispersed phased array data was used for the bursts detected in the observations on 25/02/24 and 14/03/24; phased array data was used for the other observations. All bursts were then analysed using \href{https://github.com/astrogewgaw/scarab}{\texttt{scarab}}\footnote{\url{https://github.com/astrogewgaw/scarab}}, an analysis toolkit for FRB and pulsar data being developed for the \href{https://spotlight.ncra.tifr.res.in}{\textbf{\texttt{SPOTLIGHT}}} project \footnote{The SPOTLIGHT project is a commensal survey for FRBs, RRATs, and pulsars using the uGMRT. More details can be found here: \url{https://spotlight.ncra.tifr.res.in}.}. \texttt{scarab} first dedisperses each burst to its detected DM (that is, where the burst was detected with maximum S/N from the single pulse search done previously), in order to compensate for inter-channel dispersive delays. It then corrects for the bandpass via a simple median subtraction and normalisation, and downsamples the data in frequency and time. It then clips the data to a smaller time interval centered around the burst, and masks channels where the emission from the burst is absent. The latter is done by first forming a channel mask based on whether the emission in that channel is greater than the emission in a region away from the burst (known as the \textit{off-region}). Each contiguous sub-band in this mask is then collapsed in frequency to obtain a profile, and this is convolved with a boxcar to check for the presence of a burst. Only sub-bands where the burst is present above a certain S/N threshold are used to form the final mask. While crude, this method is fast, and helps obtain a reasonable approximation to the burst's emission bandwidth. \texttt{scarab}'s S/N estimation code is based on the \href{https://bitbucket.org/vmorello/spyden}{\texttt{spyden}}\footnote{\url{https://bitbucket.org/vmorello/spyden}} package. For data recorded on 25/02/24, bursts were downsampled in time based on their S/N: bursts were downsampled to 40.96 $\mu$s, 163.84 $\mu$s, or 327.68 $\mu$s, depending on whether the S/N of the burst was $\geq$ 25, between 10 and 25, or $<$ 10, respectively. However, for data recorded on 14/03/24, bursts were downsampled to 655.36 $\mu$s, in order to compensate for the S/N loss due to using uGMRT's sub-array mode, where, since the array was divided into two frequency bands, it lead to a reduction in the gain of the PA/CD beam for each band. For the latest data recorded on 03/07/24 and 13/07/24, bursts were downsampled to 327.68 $\mu$s. All bursts were downsampled to 128 channels, which corresponds to a resolution of 1.5625 MHz.

\begin{figure}
    \includegraphics[width=0.45\textwidth]{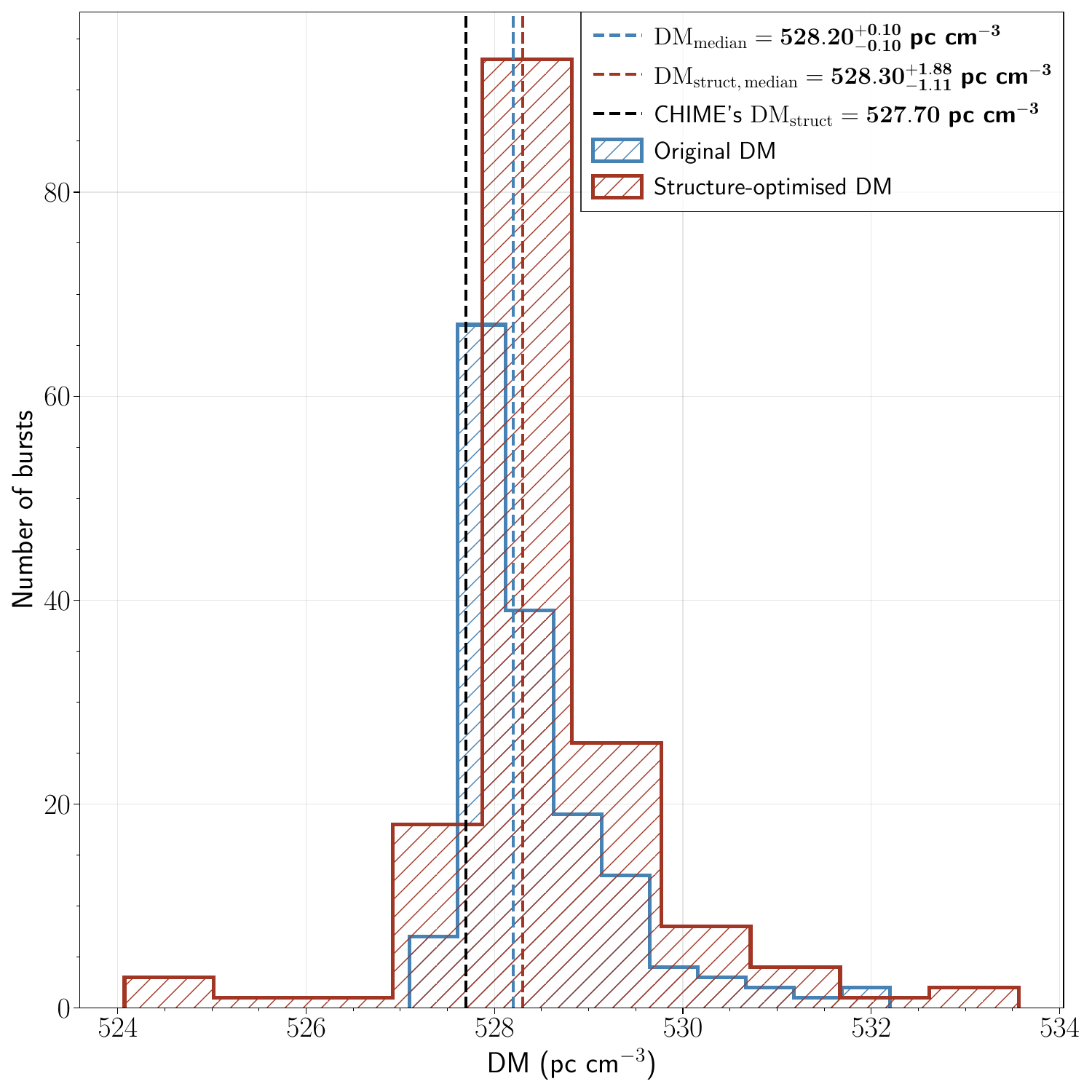}
    \caption{A comparison of the distributions of the original DM values obtained when searching for the bursts, and the structure-optimised DM values obtained using \texttt{SHRINE}'s routines in \texttt{scarab}. The brick-red line indicates the structure-optimised DM obtained by CHIME, and reported in \cite{shin_chimefrb_2024}, while the blue and orange lines correspond to the median DM values for each of the respective distributions. The median DM value remains largely unchanged.}\label{fig:dm_vs_dmopt}
\end{figure}

After the steps above, \texttt{scarab} estimates the structure-optimised DM for each burst. It uses the implementation in \href{https://github.com/marcinglowacki/SHRINE}{\texttt{SHRINE}}\footnote{\url{https://github.com/marcinglowacki/SHRINE}}, described in \cite{sutinjo_calculation_2023}. It then dedisperses each burst to its structure-optimised DM (if the estimation succeeds; otherwise, no change is made), and fits its profile and spectrum independently. For the former, it utilises the same models as those implemented in the \href{https://github.com/fjankowsk/scatfit}{\texttt{scatfit}}\footnote{\url{https://github.com/fjankowsk/scatfit}} package \citep{jankowski_sample_2023}, which were in turn taken from \cite{mckinnon_analytical_2014}. For the latter, the user can choose between a simple Gaussian model, or a running power law, as suggested and implemented in \href{https://github.com/CHIMEFRB/fitburst}{\texttt{fitburst}}\footnote{\url{https://github.com/CHIMEFRB/fitburst}} \citep{fonseca_modeling_2024}. \texttt{scarab} is also capable of fitting multiple profile components, and comes with multiple peak-finding methods.

\begin{figure}
  \begin{center}
    \includegraphics[width=0.45\textwidth]{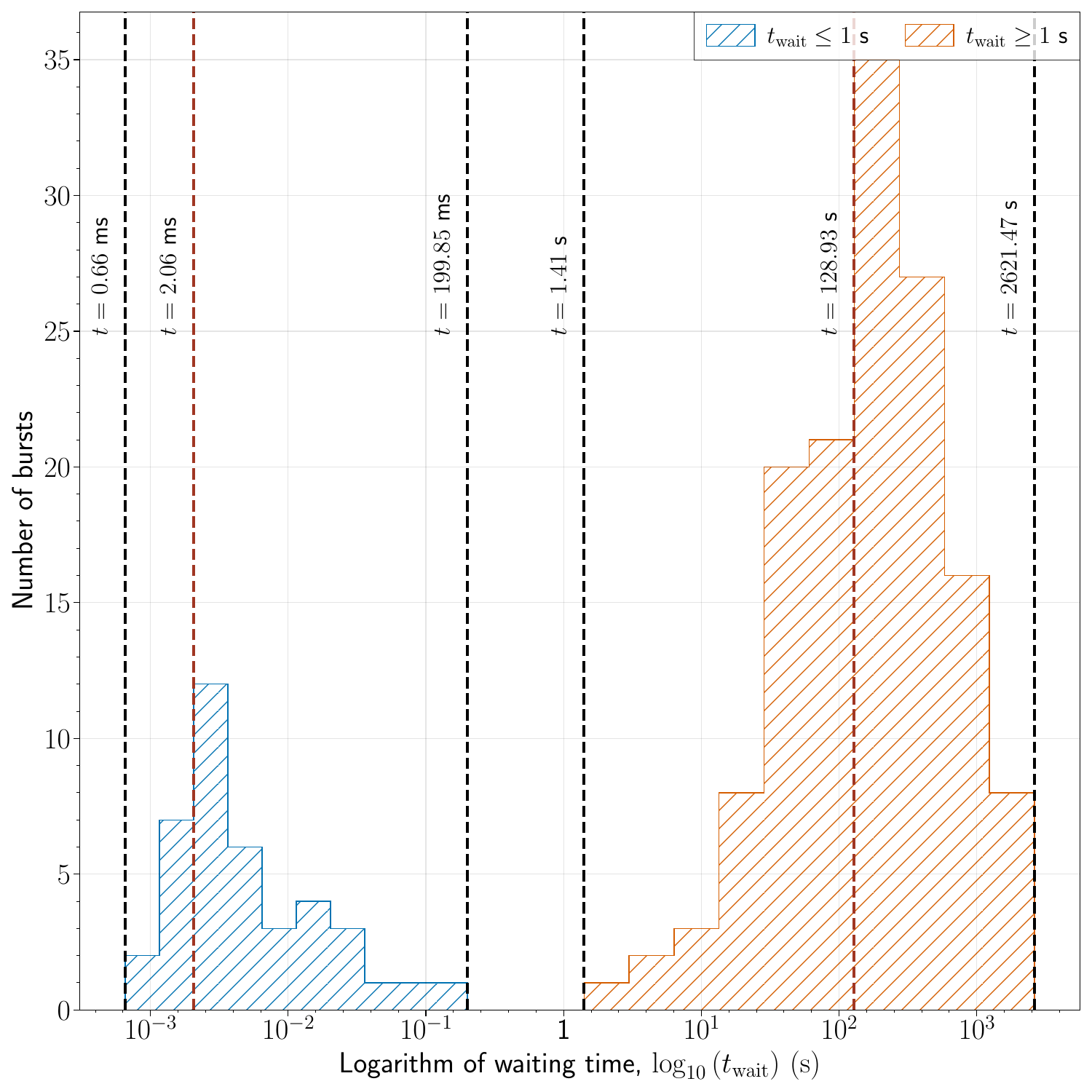}
  \end{center}
  \caption{
  A histogram of the logarithm of the waiting time, $\log_{10}\left(t_{\mathrm{wait}}\right)$; the bi-modality in waiting times is clearly visible, with a separation around $t_{\mathrm{wait}} = 1$ s.}\label{fig:waiting_time_bimodality}
\end{figure}

\begin{figure*}
  \begin{center}
    \includegraphics[width=\textwidth]{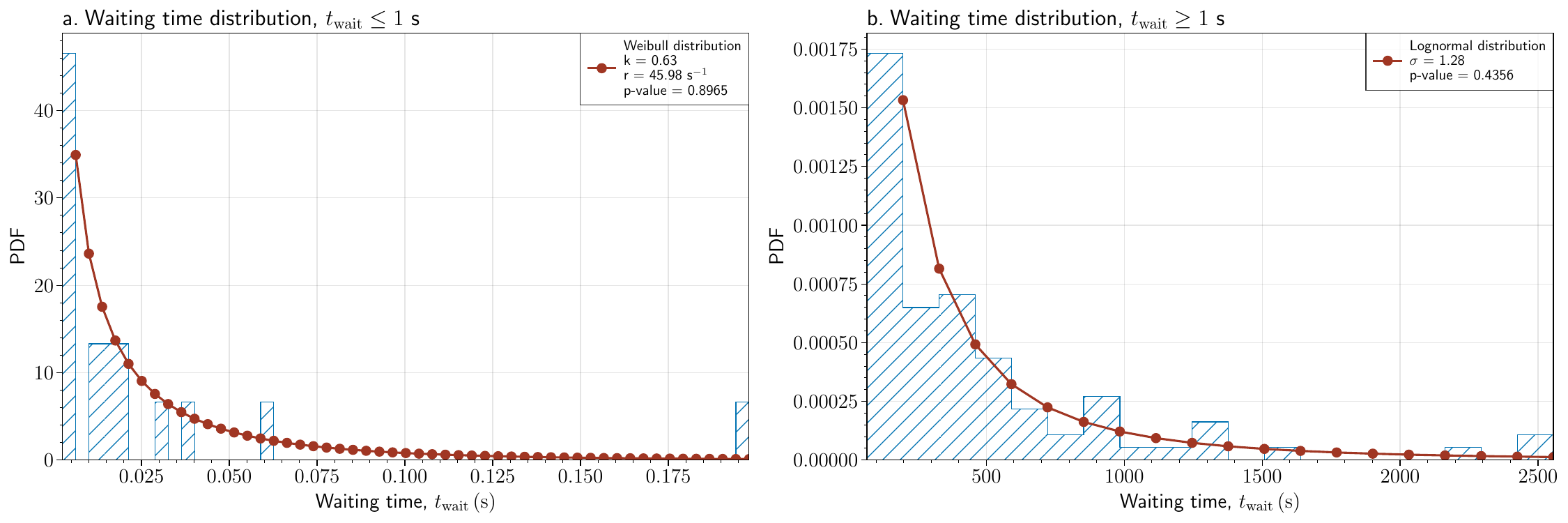}
  \end{center}
\caption{The waiting time distribution for \textbf{(a)} only for bursts with waiting times $t_{\mathrm{wait}} \leq 1$ s, and \textbf{(b)} only for bursts with waiting times $t_{\mathrm{wait}} \geq 1$ s is shown. The waiting time distribution for all bursts could not be fit with a single distribution, owing to the bi-modality readily evident from Fig. \ref{fig:waiting_time_bimodality}. Hence, the waiting time distributions for short and long waiting timescales were fit separately: the former is best fit by a Weibull distribution, while the latter is best fit by a log-normal distribution.\label{fig:waiting_time_distribution}}
\end{figure*}

After all bursts are modeled, their post-fit parameters are saved to disk, to be used for the next step: analysing burst statistics. \texttt{scarab} comes with methods to fit parametric distributions to data, based on the \href{https://github.com/erdogant/distfit/}{\texttt{distfit}}\footnote{\url{https://github.com/erdogant/distfit}} package \citep{taskesen_distfit_2020}. This allows the user to fit the values for any individual burst parameter to roughly 90 theoretical probability distributions defined in \texttt{scipy}'s \texttt{stats} module \citep{virtanen_scipy_2020}, and obtain the best fit. Each fit is then further tested using a parametric bootstrap test for the goodness-of-fit, as described in \cite{stute_bootstrap_1993, genest_validity_2008, kojadinovic_goodness--fit_2012}, and implemented in \texttt{scipy}. The test involves fitting the unknown parameters to the given data, and estimating a chosen statistic for the distribution-data pair; in our analysis, we chose the Anderson-Darling statistic. Random samples are subsequently drawn from the null-hypothesized distribution, and the unknown parameters are fitted for each sample. The chosen statistic is then computed for each fit, and its value is compared to the value obtained for the data itself. This yields a \textit{p-value} for each fit. We use the procedure described above to fit the waiting-time and isotropic energy distributions for the detected bursts.

\begin{figure*}
\gridline{\fig{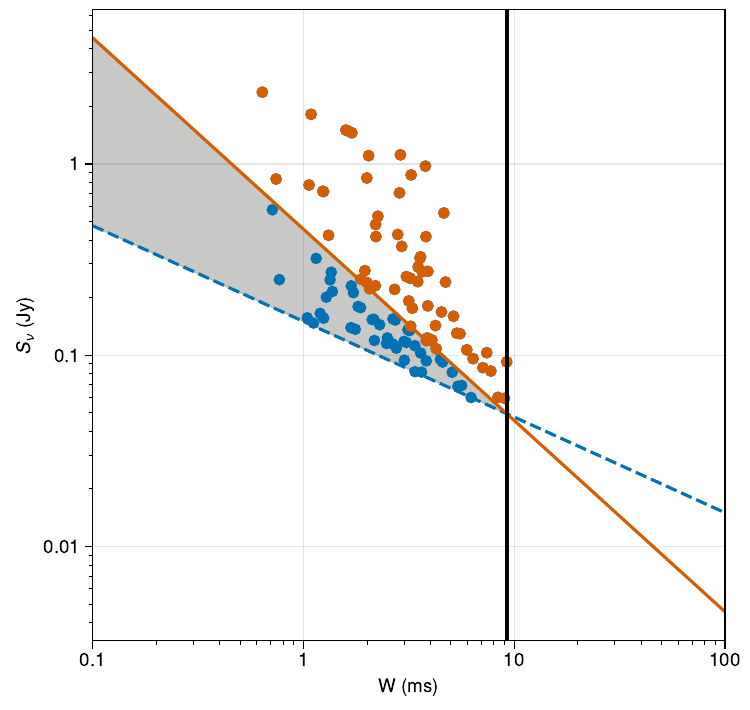}{0.45\textwidth}{(a) 25th February (Band 4)}
          \fig{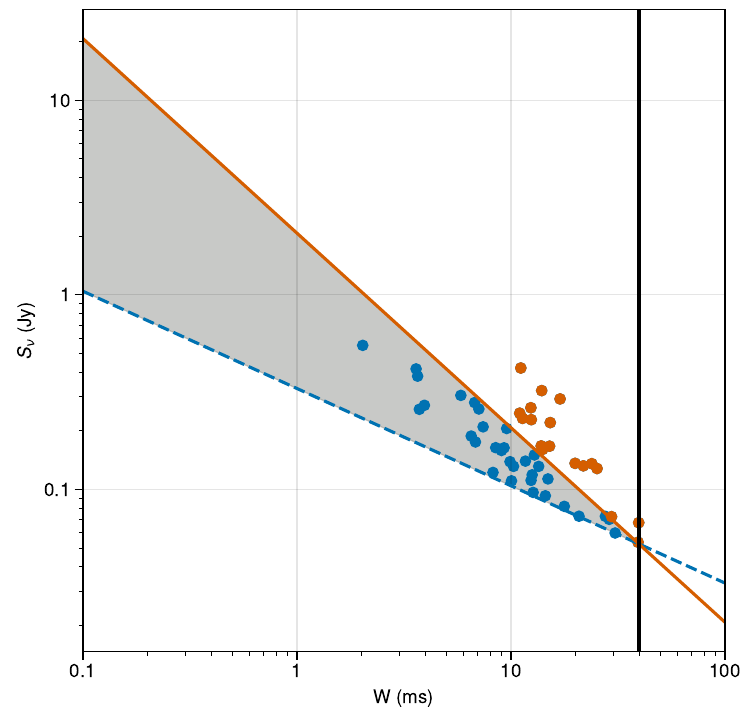}{0.45\textwidth}{(b) 14th March (Band 4)}}
\gridline{\fig{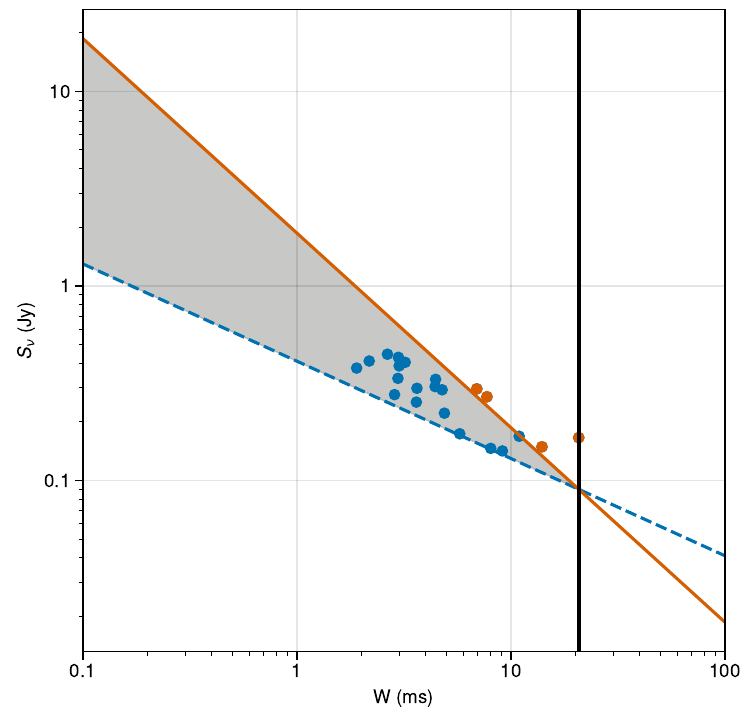}{0.45\textwidth}{(c) 14th March (Band 3)}
          \fig{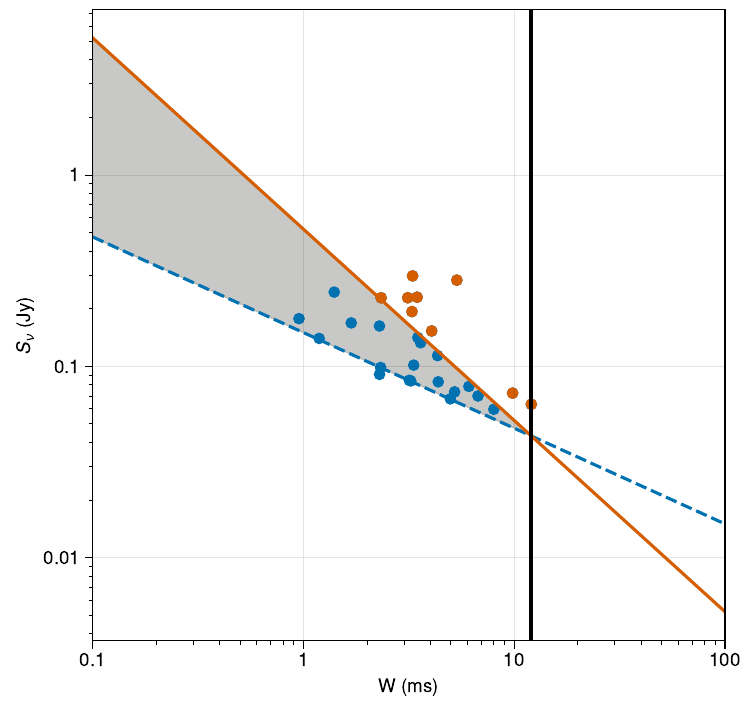}{0.45\textwidth}{(d) 3rd and 13th July (Band 4)}}
          \caption{Selecting complete bursts in each epoch, as per the methodology outlined in \cite{keane_fast_2015}. The points represent the bursts detected in each epoch, plotted in the flux$-$width plane. The dashed line represents the S/N threshold, which was the same for all epochs, and fixed to S/N = 7. The orange solid line represents the constant fluence line which intersects the S/N threshold at the maximum putative width, which is taken as the maximum width detected in each epoch. Bursts that lie in the region between these two lines (points in blue) are qualified as incomplete, while those that lie outside (points in orange) are complete.}\label{fig:completeness}
\end{figure*}

Visibility data was recorded simultaneously with the beamformed data for the observations on 25/02/24 and 14/03/24, with the source 2225$-$049 used as a bandpass, flux, and gain calibrator. The calibrator preceded every target scan. We have used 3C48 for flux calibration. An imaging analysis was carried out using this visibility data with an automated imaging pipeline, composed of \texttt{flagcal} (for flagging and calibrating GMRT visibilities) \citep{prasad_flagcalflagging_2011}, \href{https://github.com/lofar-astron/PyBDSF}{\texttt{PyBDSF}}\footnote{\url{https://github.com/lofar-astron/PyBDSF}} (for automatic source detection), and \texttt{CASA} \citep{team_casa_2022} for deconvolution and self-calibration. Two loops of phase-only, and final amplitude-phase calibration were carried out for self-calibration.

\section{Results}\label{section:results}

\begin{figure}
  \begin{center}
    \includegraphics[width=0.45\textwidth]{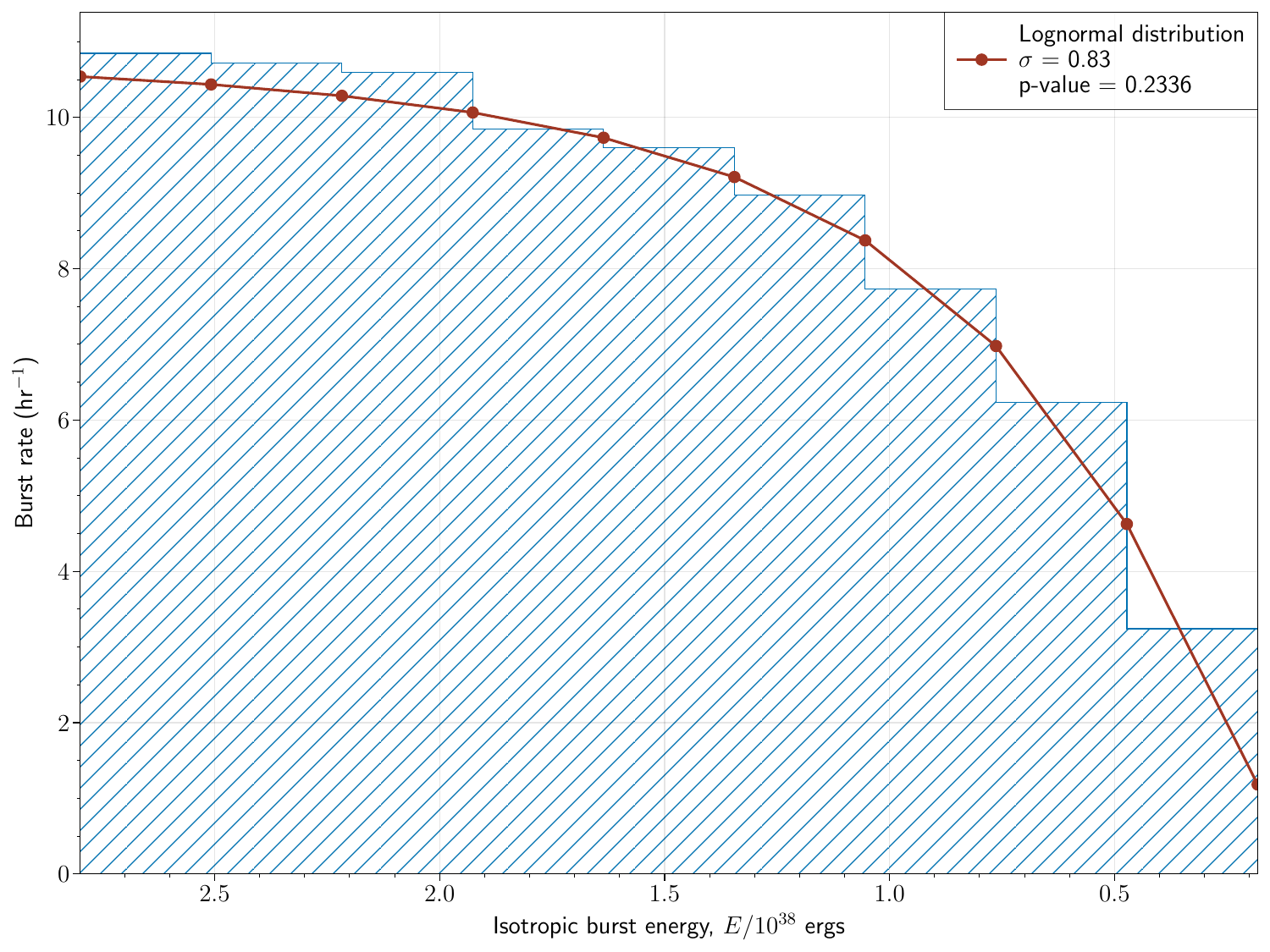}
  \end{center}
  \caption{The burst rate v/s isotropic energy distribution of the detected bursts, overlaid by the best fit, a log-normal distribution with $\sigma = 0.83$. Only complete bursts were used to create the distribution, with completeness taken into account on an epoch-by-epoch basis, using the methodology outlined in \cite{keane_fast_2015}.}\label{fig:energy_distribution}
\end{figure}

We detected a total of 167 bursts from our observations\footnote{Refer to the GitHub repository of this paper: \url{https://github.com/astrogewgaw/240114A}, for the plots of all detected bursts, and their post-fit parameters. The latter are stored as CSV files by epoch, enabling easier inclusion into other analyses. The repository also includes high-resolution versions of the plots in this paper, as well as the paper's source code.}. Fig \ref{fig:gallery} shows 9 of the brightest bursts detected from our data, showcasing the wide variety of morphological features shown by the bursts, including band-limited emission, multiple profile components, and sub-burst frequency drifting (that is, the \textit{sad trombone effect}). The intrinsic widths of the bursts varied from 0.246 to 39.364 ms, while the scattering timescales varied from 0.004 to 28.289 ms. The band occupancy varied from approximately 9 to 180 MHz; note that the former covers only 4.5\% of the band, while the later covers almost 90\% of it, indicating that the band occupancy of the bursts varied widely. 56\% of the bursts had a band occupancy equal to or below 50\%, indicating that a majority of them were extremely band-limited.

Since the source was observed on 4 different dates, it allowed us to determine how its activity and properties changed over time. Fig. \ref{fig:distributions} shows the distributions of the structure-optimised $\mathrm{DM}$, the intrinsic width, $W_{\mathrm{int}}$, the scattering width, $W_{\mathrm{scat}}$, and the flux of the detected bursts, plotted as violin plots, on each of the 4 separate observation dates. From the figure, it is evident that the bursts detected on 14th March were wider and more scattered on average compared to the bursts detected on either 25th February, or 3rd and 13th July. These bursts also showed a larger variation in their DM values, even after structure-optimisation. The fluxes of the bursts seem to be gradually decreasing over time as well. The burst rate changed substantially over our observation campaign (see Table \ref{table:obsdetails}), indicating that over time the emission from FRB 20240114A is evolving.

In the following sub-sections, we discuss the statistical distribution of a few burst properties in detail; we cover the DM distribution in a bit more detail in \S\ref{subsection:dmdists}, the waiting time distributions in \S\ref{subsection:wtdists}, and the isotropic energy distribution in \S\ref{subsection:edists}. We finally wrap up with a brief discussion about the persistent radio source (PRS) associated with this FRB in \S\ref{subsection:prs}.

\begin{deluxetable*}{ccllr}
\tablecaption{Waiting time and energy distribution fits for various hyperactive FRBs\label{table:comparison}}
\tablehead{\colhead{FRB} & \colhead{$N$} & \colhead{Waiting time} & \colhead{Energy} & \colhead{}}
\startdata
\multirow{5}{*}{20121102A}     &     28 & Weibull, $k = 0.34^{+0.06}_{-0.05}$ & $-$ & \cite{oppermann_non-poissonian_2018} \\
                               &    157 & $-$ & Power law, $\alpha = -1.6$ to $-1.8$ & \cite{wang_universal_2019} \\
                               &     41 & Weibull, $k = 0.62^{+0.1}_{-0.09}$ & Power law, $\alpha = -1.1 \pm 0.2$ & \cite{cruces_repeating_2021} \\
                               &        & Weibull, $k = 0.73^{+0.12}_{-0.10}$ ($\delta t > 1$s) & & \\
                               &   1652 & Log-normal, $\sigma = 1.27 \pm 0.02$ & Log-normal + Cauchy, $\sigma = 0.52$, $\alpha = 1.85 \pm 0.3$ & \cite{li_bimodal_2021} \\
                               &    133 & Log-normal, $\sigma = 1.12 \pm 0.07$ & Broken power law, $\alpha_{1} = -0.4 \pm 0.1$, $\alpha_{2} = -1.8 \pm 0.2$ & \cite{aggarwal_comprehensive_2021} \\
                               &    478 & Log-normal, $\sigma = 1.22 \pm 0.04$ & Broken power law, $\alpha_{1} = -1.38 \pm 0.01$, $\alpha_{2} = -1.04 \pm 0.02$ & \cite{hewitt_arecibo_2022} \\
\tableline
\multirow{2}{*}{20180916B}     &    195 & Log-normal & $-$ & \cite{wang_statistical_2023} \\
                               &     78 & Weibull, $k = 0.84 \pm 0.02$ & $-$ & \cite{bhattacharyya_wideband_2024} \\
\tableline
20190520B                      &     79 & Weibull, $k = 0.37 \pm 0.04$ & Log-normal & \cite{niu_repeating_2022} \\
                               &        & Weibull, $k = 0.76^{+0.09}_{-0.08}$ ($\delta t > 1$s) & & \\
\tableline
\multirow{3}{*}{20201124A}     &  1,863 & Log-normal & Broken power law,  $\alpha_{1} = -0.36 \pm 0.02$, $\alpha_{2} = -1.5 \pm 0.1$ & \cite{xu_fast_2022} \\
                               & $>$800 & Log-normal, $\sigma = 1.33 \pm 0.03$ & Broken power law, $\alpha_{1} = -1.22 \pm 0.01$, $\alpha_{2} = -4.27 \pm 0.23$ & \cite{zhang_fast_2022}\\
                               &    996 & Log-normal, $\sigma = 1.04 \pm 0.03$ & $-$                                                 & \cite{niu_fast_2022} \\
\tableline
\multirow{2}{*}{20220912B}     &    696 & Weibull, $k = 0.88 \pm 0.01$ ($\delta t > 1$s) & Power law, $\alpha = -1.67 \pm 0.01$ & \cite{konijn_nancay_2024} \\
                               &   1076 & Log-normal & Broken power law, $\alpha_{1} = -0.38 \pm 0.02$  $\alpha_{2} = -2.07 \pm 0.07$ & \cite{zhang_fast_2023} \\
\tableline
20200120E                      &     53 & Weibull, $k = 0.50^{+0.04}_{-0.05}, 0.69^{+0.20}_{-0.17}$ & Power law, $\alpha = -2.39 \pm 0.12$ & \cite{nimmo_burst_2023} \\
\tableline
\multirow{2}{*}{20240114A}     &    167 & Weibull, $k = 0.63$ $(\delta t < 1 \text{ s})$ & Log-normal, $\sigma = 0.83$ & This work. \\
                               &        & Log-normal, $\sigma = 1.28$ $(\delta t > 1 \text{ s})$ &  &  \\
\enddata
\end{deluxetable*}

\subsection{DM Distribution}\label{subsection:dmdists}

Fig. \ref{fig:dm_vs_dmopt}
shows the comparison between distributions of the original DM values (obtained via a DM search), and the structure-optimised DM values obtained using \texttt{scarab}, which internally uses the algorithm behind \texttt{SHRINE}. Compared to other implementations, such as \texttt{DM\_PHASE} \citep{seymour_dm_phase_2019} or \texttt{DM\_POWER} \citep{lin_dm-power_2023}, \texttt{SHRINE} has a few added advantages, such as being able to provide a direct error estimate for each structure-optimised DM, and being able to carry out the estimation for low S/N bursts. \texttt{SHRINE} was unable to estimate the structure-optimised DM for only 10 bursts, mostly due to high amounts of RFI in their data. The structure-optimised DM of the bursts varies from 524.07 to 533.56 pc cm$^{-3}$, with a median value of $528.36^{+1.88}_{-1.11}$ pc cm$^{-3}$.

\subsection{Waiting Time Distribution}\label{subsection:wtdists}

The waiting time, $t_{\mathrm{wait}}$, was fit for all observations, and for two subsets: $t_{\mathrm{wait}} \geq 1$ s, and $t_{\mathrm{wait}} \leq 1$ s. This was done since a clear bi-modality could be seen in the distribution at $t_{\mathrm{wait}} = 1$ s (see Fig \ref{fig:waiting_time_bimodality}). Different observations are treated as independent when estimating the waiting time; as demonstrated by \cite{oppermann_non-poissonian_2018}, the error introduced by this approximation is minor. The exponential, Weibull, log-normal, and gamma distributions were tried for all fits. None of the distributions fit the waiting time distribution for all observations; in other words, the p-values obtained were $\approx 0.0001$, which means that we can reject the null hypothesis, that the data was drawn from any of the distributions above, at a $99.99\%$ confidence level. On the other hand, when the waiting time distributions for the two subsets were fit separately, they yielded better results: the waiting time distribution for $t_{\mathrm{wait}} \leq 1$ s was best fit by a Weibull distribution with a p-value of 0.8965, while the waiting time distribution for $t_{\mathrm{wait}} \geq 1$ s was best fit by a log-normal distribution, with a p-value of 0.4356. Note that the smaller p-value for the latter case indicates a worse fit, but was the best among all the distributions tried, and does not allow us to reject the null hypothesis at any common significance level. The plots and best fits for the waiting time distributions for the two subsets can be seen in Fig. \ref{fig:waiting_time_distribution}. The waiting time distribution was best fit by a log-normal distribution, with $\sigma = 1.28$, while the removed bursts follow a Weibull distribution as expected, with a shape parameter $k = 0.63$, and a corresponding mean rate of $r = 45.98 \text{ s}^{-1}$. This implies that a majority of the clustering can be attributed to subcomponents being treated as independent bursts.

\cite{katz_log-normal_2024} posits that the width of the log-normal waiting time distribution of transient phenomena functions as a critical exponent, and can be used to divide them into universality classes, in a manner similar to the theory of critical phenomena \citep{pelissetto_critical_2002}. For instance, perfectly periodic phenomena, such as pulsars, have $\sigma \approx 0$. On the other hand, shot noise --- that is, a process following Poissonian statistics --- will have $\sigma = 0.723$; a detailed calculation can be found in \cite{katz_log-normal_2024}. The value of $\sigma$ of phenomena in the same class are not necessarily required to be close, but rather can be different up to $\ll 1$.

Keeping the above in mind, we note that the waiting distribution of several FRBs are known to be well-fit by a log-normal distribution; several examples can be found in Table \ref{table:comparison}. The value of $\sigma = 1.28$ obtained through our analysis is close to the values for FRB 20121102A ($\sigma = 1.27, 1.12, 1.22$), and FRB 20201124A ($\sigma = 1.33, 1.04$), indicating that their activity is distributed similarly across time, and could be the result of the same, or similar, emission mechanism(s). Values of $\sigma$ greater than the one for shot noise, $\sigma = 0.723$, indicate varying rates of activity, since shot noise has the most random possible statistics if the mean or statistically expected rate of activity is unchanging. The fact that the value of $\sigma$ we estimate is $\sim 1.8$ times larger indicates that the activity of FRB 20240114A is changing, and that its emission mechanism may have some long term memory, as has been previously indicated for other FRBs as well (for instance, \cite{wang_memory_2024}).

\subsection{Isotropic Energy Distribution}\label{subsection:edists}

The flux for each burst was estimated using the \href{https://github.com/astrogewgaw/gmrtetc}{\texttt{gmrtetc}}\footnote{\url{https://github.com/astrogewgaw/gmrtetc}} package, which is a pure Python implementation of GMRT's Exposure Time Calculator. \texttt{gmrtetc} gives an estimate of the RMS for each burst detected in a particular epoch, based on observational parameters, and the measured intrinsic and scattering width of the burst\footnote{For a more detailed overview of how GMRT's Exposure Time Calculator functions, refer to the help document available here: \url{http://www.ncra.tifr.res.in:8081/~secr-ops/etc/etc_help.pdf}}. These estimates are then scaled by the detection S/N to obtain the flux density. The fluence of each burst is then simply a product of the flux density and the effective width, $W_{\mathrm{eff}} = \sqrt{W_{\mathrm{int}}^{2} + W_{\mathrm{scat}}^{2}}$, where $W_{\mathrm{int}}$ is the intrinsic width, and $W_{\mathrm{scat}}$ is the scattering width. Note the S/N does not take into account all components of a bursts; it only takes the brightest component into account, if the burst has multiple components. Using the estimated flux and fluence values, the completeness threshold was estimated via the methodology outlined in \cite{keane_fast_2015}, taking an S/N threshold of 7.0, and setting the putative width to the maximum width detected in each epoch (see Fig. \ref{fig:completeness}). After the bursts below the completeness threshold were removed, a set of 91 bursts was obtained. Then, the isotropic energy released from each burst was calculated using \citep{aggarwal_observational_2021}:

\begin{equation}
  E =
    4 \pi \times
    {\left(\frac{D_{L}}{\mathrm{cm}}\right)}^{2}
    \left(\frac{F}{\mathrm{Jy}\,{\rm{s}}}\right)
    \left(\frac{\Delta \nu_{\mathrm{occ}}}{\mathrm{Hz}}\right)
    \times 10^{-23} \,
    \mathrm{erg},
  \label{eq:energycalc}
\end{equation}

where $D_{L} = 630.72 \text{ Mpc}$ is the luminosity distance of FRB 20240114A (for a redshift of $z = 0.13$ \citep{bhardwaj_redshift_2024}, and using the cosmological parameters given in \cite{aghanim_planck_2020} by the Planck Collaboration), $F$ is the fluence, $\Delta \nu_{\mathrm{occ}}$ is the band occupancy, and E is the isotropic energy of each burst. After calculating the isotropic energy of each burst using Eq. \ref{eq:energycalc}, we fit the selected distributions to the isotropic energies for these bursts, using the approach described in \S\ref{section:analysis}. The log-normal, exponential, power law, and Cauchy distributions were tried on the data, and a best fit was obtained using the log-normal distribution, with $\sigma = 0.83$, and a p-value $= 0.2336$. The low p-value, while the best among the distributions tried, indicates that the log-normal distribution may not be a good fit to the data. It is high enough, however, to not allow us to reject the null hypothesis at any common significance level. The corresponding cumulative distribution of burst rate v/s isotropic energy is shown in Fig. \ref{fig:energy_distribution}.

\subsection{Persistent Radio Source}\label{subsection:prs}

To date, 3 FRBs have been associated with a persistent radio source (PRS): FRB 20121102A \citep{chatterjee_direct_2017}, FRB 20190520B \citep{niu_repeating_2022}, and FRB 20201124A \citep{bruni_nebular_2024}. From Table \ref{table:comparison}, it can be seen that all of these FRBs have shown periods of hyperactivity. Thus, a similar detection was expected for FRB 20240114A, given its well-established hyperactive nature. However, a search for such a source in archival radio survey data by the authors of \cite{bhardwaj_redshift_2024} yielded nothing, and they placed a 300 $\mu$Jy upper limit using FIRST survey data, taken at 1.5 GHz. A more stringent upper limit had already been placed, using observation in uGMRT's Band 4 (550 to 750 MHz), by \cite{kumar_detection_2024}. From our previous analysis, we had also placed an upper limit of 124 $\mu$Jy with a 5$\sigma$ significance, in the absence of a detection. However, using data from MeerKAT observations at L-band, \citep{zhang_discovery_2024} first associated a PRS with FRB 20240114A, and estimated a flux density of $72 \pm 14$ $\mu$Jy. This prompted us to revisit our data, and we were able to obtain a detection at 3$\sigma$ for the PRS, with an estimated flux density of $70 \pm 22$ $\mu$Jy, at uGMRT's Band 4 (550 to 750 MHz). Using the same band, and a deeper scan, \citep{bhusare_detection_2024} obtained an 8$\sigma$ detection, with a peak flux density of $65.6 \pm 8.1$ $\mu$Jy. Earlier observations by \cite{bruni_discovery_2024} using the VLBA at 5 GHz confirm the presence of the PRS, with an estimated flux density of $46 \pm 9$ $\mu$Jy. The presence of a PRS, the association with a star-forming dwarf galaxy, and periods of hyperactivity seen for FRB 20240114A have also been seen for two other FRBs: FRB 20121102A, and FRB 20190520B.

\section{Summary}\label{section:summary}

We detected a total of 167 bursts from FRB 20240114A using the uGMRT. Through our analysis of the statistics of the detected bursts, we obtained the following results:

\begin{itemize}
  \item Majority of the bursts were band-limited, with 56\% of the bursts with a band occupancy of 50\% or less. Band occupancies varied widely, from 9 MHz (4.5\%) to 180 MHz ($\sim 90$\%).
  \item The bursts detected on 14th March were wider and more scattered than bursts detected in other epochs (see Fig. \ref{fig:distributions}). They also show a wider variation in their dispersion measures.
  \item The fluxes of the detected bursts decreased gradually over time (see Fig. \ref{fig:distributions}). The burst rate changed dramatically as well over the course of the observation campaign, indicating a siginificant evolution in the activity of the source.
  \item The waiting time distribution displayed a clear bi-modality, and hence the waiting times were split into two subsets. The subset with $t_{\mathrm{wait}} > 1$s was better fit by a log-normal distribution, with $\sigma = 1.28$, while the subset with $t_{\mathrm{wait}} < 1$s was best fit by a Weibull distribution, with $k = 0.63$, and a mean rate of $r = 45.98 \text{ s}^{-1}$. This implies that a majority of the clustering can be attributed to subcomponents being treated as independent bursts.
  \item The cumulative burst rate v/s isotropic energy distribution was best fit by a log-normal distribution, with $\sigma = 0.83$.
  \item A 3$\sigma$ detection for the PRS associated with FRB 20240114A was obtained from our data, with a estimated flux density of $70 \pm 22$ $\mu$Jy, at uGMRT's Band 4 (550 to 750 MHz).
\end{itemize}

\begin{acknowledgments}

We would like to thank our anonymous reviewer for the insightful comments, which helped greatly improve the paper's content. We also gratefully acknowledge the Department of Atomic Energy, Government of India, for its assistance under project No. 12-R\&D-TFR-5.02-0700. Furthermore, we are grateful to the GMRT Operations Team for the prompt approval and scheduling of the DDT observations that allowed us to probe this FRB while it was in its active state, and the staff of the GMRT that made these observations possible. GMRT is run by the National Centre for Radio Astrophysics of the Tata Institute of Fundamental Research.

\end{acknowledgments}

\bibliography{refs}{}
\bibliographystyle{aasjournal}

\end{document}